\documentclass[twocolumn,english]{revtex4}
\usepackage[T1]{fontenc}
\usepackage[latin1]{inputenc}
\usepackage{graphicx}

\makeatletter

\usepackage{babel}
\makeatother
\begin{document}

\title{Effective medium theory of negative index composite metamaterials}

\author{A.I. C\u{a}buz, D. Felbacq and D. Cassagne}

\affiliation{Groupe d'Etude des Semiconducteurs UMR CNRS-UM2 5650, CC074, Université
Montpellier II, Pl. E. Bataillon, F-34095 Montpellier Cedex 05, France}

\begin{abstract}
In the homogenization of composite metamaterials the role played by
the relative positions of the wires and resonators is not well understood,
though essential. We present a general argument which shows that the
homogenization of such metamaterials can be seen as the homogenization
of 1D single negative stacks. The sensitivity to the geometrical parameters
is due to the fact that light sees the harmonic mean of the dielectric
constant when propagating parallel to the layers. Since the dielectric
constant is not positive definite the harmonic mean is \emph{not bounded}
and presents abrupt variations. We discuss applications of this remarkable
phenomenon.
\end{abstract}
\maketitle
Meta-materials with a possibly negative index of refraction have been
under intense study since the results of Pendry \cite{Pendry1998,Pendry1999}
suggested that they might be fabricated by interspersing metallic
wires (giving an effective negative permittivity) with magnetic resonators
such as split rings or dielectric fibers (giving an effective negative
permeability). However the details of the exact way in which these
two effects mix and perturb each other are not well understood. In
this letter we present an effective medium theory that accounts quantitatively
for this behavior, provides an explicit geometrical recipe for obtaining
true negative index media and opens the way to new applications.

No general argument exists that by mixing magnetic resonators and
wires one must obtain a negative index material at some frequency.
Indeed, as the results of Pokrovsky and Efros have shown \cite{Pokrovsky2002},
the negative epsilon effect of metallic gratings is fragile with respect
to the brutal mixing of the metallic wires with other objects. Negative
epsilon is essentially an interference phenomenon and anything that
perturbs the interference of the waves scattered by neighboring wires
is liable to destroy it. The wires need to {}``see'' each other.
The results of Pokrovsky and Efros are the first indication that the
effective medium depends critically on the microscopic geometry \cite{Marques2004a}.
They essentially tell us that resonators should not be placed in between
the wires, or the aggregate may lose its negative permittivity. The
regions occupied by the wires must have higher connectivity, and the
simplest way of acheiving this is a stack of alternating rows of wires
and resonators. Each row of wires will then act as a homogeneous slab
with negative permittivity ($\varepsilon=1-\frac{\omega_{ep}^{2}}{\omega^{2}}$
and $\omega_{ep}=\frac{2\pi c_{0}^{2}}{a^{2}\ln\left(a/r\right)}$
\cite{Pendry1998,Felbacq2005a}), and each row of resonators will
act as a homogeneous slab with negative permeability ($\mu=1-\frac{a\omega^{2}}{\omega^{2}-\omega_{mp}^{2}}$,
with $a$ and $\omega_{mp}$ determined by the geometry \cite{Pendry1999,Felbacq2005}).
We thereby arrive at the remarkable and quite general conclusion that
the homogenization of complex wire/resonator composites is the homogenization
of 1D stacks of alternating single negative homogeneous slabs. This
conclusion is a direct consequence of the fact that the negative permittivity
of wire gratings depends critically on their mutual interferences,
in contrast to the negative permeability effect which is essentially
a resonance phenomenon and does not depend critically on the coupling
between neighboring resonators \cite{Marques2002a,Felbacq2005a}.

However, while some interesting results on 1D single negative media
have been published \cite{Alu2003,Yoon2005,Kim2005}, the natural
question of the precise conditions under which such structures can
be homogenized to obtain a negative index medium has not been satisfactorily
answered. 

\begin{figure}
\begin{center}\includegraphics[%
  scale=0.45]{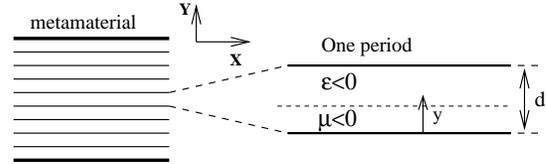}\end{center}

\caption{\label{cap:unitcellmetamaterial}The structure we study is a 1D periodic
medium with a unit cell composed of two layers, one with negative
$\varepsilon$ (and positive $\mu$), one with negative $\mu$ (and
positive $\varepsilon$). The $\mathbf{H}$ field is directed in the
Z direction and the incident field is directed upwards.}
\end{figure}

In this work we offer a rigorous self-contained homogenization analysis
that gives a clear description of the 1D alternating single negative
medium described above and pictured schematically in Figure 1. We
shall use the term {}``$P\cdot k$ negative'' to unambiguously refer
to waves or homogeneous materials that may support waves which have
a dot product of the Poynting vector and wave vector that is negative.

The discriminating factor, as in any homogenization problem, is the
ratio between the wavelength and the period of the structure. In our
case a complication is introduced by the extreme dispersion of resonator
behavior, but we avoid it by fixing the wavelength (and therefore
the internal resonator geometry) and scaling the \emph{relative} \emph{placement}
of wires and resonators. The discriminating geometrical parameter
of the homogeneous model is therefore the ratio $r=\frac{\lambda}{d}$
where $\lambda$ is the free space wavelength and $d$ is the period.
This parameter is, a priori, large, which authorizes us to use the
well-established \emph{two-scale expansion} approach. In the following
we treat only the $\mathbf{H}$ polarization case. The results hold
in $\mathbf{E}$ polarization as well, provided we exchange $\mu$
and $\varepsilon$ everywhere. 

We note the macroscopic variables with capital letters $X$ and $Y$
and the microscopic variable with the small letter $y$ as in Fig.
1. We first need to introduce the small parameter $\eta=d/\lambda$
which will be used in the expansion. Next we define the microscopic
variable $y=\frac{Y}{\eta}$ which varies fast, $\eta$ being small,
so it describes the system on the microscopic scale, the scale of
an individual layer. Since we are interested in the homogeneous behavior
of the system, at the end of the analysis this variable will be averaged
over and will disappear. It can be regarded as an internal degree
of freedom of the unit cell.

For a 1D structure with piecewise continuous $\varepsilon(Y)$ and
$\mu(Y)$ Maxwell's equations can be written as\begin{equation}
\frac{d}{dY}\left(\frac{1}{\varepsilon(Y)}\frac{dU}{dY}\right)+\left(k^{2}\mu(Y)-\frac{k^{2}\sin^{2}(\theta)}{\varepsilon(Y)}\right)U(Y)=0\label{eq:mastereq}\end{equation}
 where $H(Y)=U(Y,y)\textrm{e}^{ikx\sin(\theta)}$, $k=2\pi/\lambda$,
$\theta$ is the angle of incidence in the $z=0$ plane and where
we will note $k^{2}\textrm{sin}^{2}(\theta)=k_{X}$ below. This second
order differential equation can be written as a system of two first
order equations \begin{equation}
\xi=\frac{1}{\varepsilon}\frac{dU}{dY}\label{eq:xidefinition}\end{equation}
\begin{equation}
\frac{d\xi}{dY}+\left(k^{2}\mu-\frac{k_{X}^{2}}{\varepsilon}\right)U=0\label{eq:mastereqchgvar}\end{equation}
 where we have introduced the new variable $\xi$ defined by equation
\ref{eq:xidefinition} and where special care must be taken
with the derivative operator as it now takes the form \begin{equation}
\frac{d}{dY}=\frac{\partial}{\partial Y}+\frac{1}{\eta}\frac{\partial}{\partial y}.\label{eq:dYoperator}\end{equation}

We develop the quantities of interest in multi-scale series with parameter
$\eta$ \begin{equation}
\begin{array}{cc}
U(Y)=U_{0}(Y,y)+\eta U_{1}(Y,y)+\cdots\\
\xi(Y)=\xi_{0}(Y,y)+\eta\xi_{1}(Y,y)+\cdots\end{array}\label{eq:multiscale}\end{equation}
 and insert them into equations~\ref{eq:xidefinition} and
\ref{eq:mastereqchgvar}. Note that all quantities on the right
side of the equations must be periodic in $y$ with period $d$. By
identifying same order terms in $\eta$ we obtain the following relations.
From Eq.~\ref{eq:xidefinition} we obtain \begin{eqnarray}
\frac{\partial U_{0}}{\partial y}=0 & \textrm{and} & \varepsilon\xi_{0}=\frac{\partial U_{0}}{\partial Y}+\frac{\partial U_{1}}{\partial y}\label{eq:xibeforeaverage}\end{eqnarray}
 while from Eq.~\ref{eq:mastereqchgvar} we obtain \begin{eqnarray}
\frac{\partial\xi_{0}}{\partial y}=0 & \textrm{and} & \frac{\partial\xi_{0}}{\partial Y}+\frac{\partial\xi_{1}}{\partial y}+\left(k^{2}\mu-\frac{k_{X}^{2}}{\varepsilon}\right)U_{0}=0.\label{eq:Ubeforeaverage}\end{eqnarray}

We notice that the zeroth order terms of the $U$ and $\xi$ expansions
do not depend on $y$, the microscopic variable. However the first
order terms do depend on it and are periodic of period $d$ as noted
above. This must be kept in mind as we proceed to average the equations
(to obtain the homogeneous behavior of the fields) because the derivative
of any continuous periodic function averages to zero. We note the
average over $y$ of a quantity $f$ as $\left\langle f\right\rangle =\frac{1}{d}\int_{0}^{d}f(y)\textrm{d}y$. 

Taking the average of \ref{eq:xibeforeaverage} and \ref{eq:Ubeforeaverage}
the $y$ derivatives disappear as explained above and we get\[
\left\langle \varepsilon\right\rangle \xi_{0}=\frac{\partial U_{0}}{\partial Y}\]
\begin{equation}
\frac{\partial\xi_{0}}{\partial Y}+\left(k^{2}\left\langle \mu\right\rangle -\left\langle \frac{1}{\varepsilon}\right\rangle k_{X}^{2}\right)U_{0}=0.\label{eq:homogenmastereq}\end{equation}
 which we put back together into one second order equation \[
\frac{\partial^{2}U_{0}}{\partial Y^{2}}+\left(k^{2}\left\langle \varepsilon\right\rangle \left\langle \mu\right\rangle -\frac{\left\langle \varepsilon\right\rangle }{\left\langle \varepsilon^{-1}\right\rangle ^{-1}}k_{X}^{2}\right)U_{0}=0.\]

The quantity in parentheses is the propagation constant in the $Y$
direction, $k_{Y}^{2}$ and we thus obtain the dispersion relation
of the homogeneous medium:

\begin{equation}
\frac{k_{X}^{2}}{\varepsilon_{Y}}+\frac{k_{Y}^{2}}{\varepsilon_{X}}=k^{2}\mu_{\textrm{eff}}\,\,\,\,\,\,\,\,\textrm{where}\label{eq:disprel}\end{equation}

\begin{equation}
\mu_{\textrm{eff}}=\left\langle \mu\right\rangle ,\,\,\varepsilon_{X}=\left\langle \varepsilon\right\rangle ,\,\,\varepsilon_{Y}=\left\langle \varepsilon^{-1}\right\rangle ^{-1}.\label{eq:permittivity_anis}\end{equation}

The relations \ref{eq:permittivity_anis} summarize the effective
medium model and the rest of the letter is concerned with exploring
their consequences. The most important of them are: 1). The medium
is anisotropic, 2). Its anisotropy can be extreme due to the unboundedness
of the harmonic mean, 3). This behavior is strongly dependent on the
microscopic geometry of the composite. We offer explicit expressions
illustrating the third point below while the first two are discussed
after the presentation of the numerical results validating our model.
We should note that we have not made any assumption about the losses
so, in general, all quantities in \ref{eq:permittivity_anis}
are complex. 

Eqs.~\ref{eq:disprel} and \ref{eq:permittivity_anis}
show that for $\mathbf{H}$ polarization, light sees the arithmetic
mean of epsilon when propagating in the $Y$ direction but the harmonic
mean when propagating in the $X$ direction. Let us denote the thicknesses
of the two layers $h_{1}$ and $h_{2}$ (we assume $h_{1}+h_{2}=1$
for simplicity) and the corresponding permittivities $\varepsilon_{1}=a_{1}+ib_{1}$
and $\varepsilon_{2}=a_{2}+ib_{2}$. Rewriting Eq. \ref{eq:permittivity_anis}
we have:\begin{eqnarray}
 & \varepsilon_{X}=h_{1}\varepsilon_{1}+h_{2}\varepsilon_{2},\,\,\, & \varepsilon_{Y}=\frac{\varepsilon_{1}\varepsilon_{2}}{h_{1}\varepsilon_{2}+h_{2}\varepsilon_{1}}\label{eq:relsanis_detail}\end{eqnarray}
 In this form it is easiest to see how the three main consequences
come about. The first one is obtained by setting $\varepsilon_{X}=\varepsilon_{Y}$.
One quickly arrives at $\varepsilon_{1}=\varepsilon_{2}$ which contradicts
our assumptions. The second one is seen by noting that there is nothing
preventing the denominator of $\varepsilon_{Y}$ from approaching
zero, ignoring losses. The third results from the prominent role played
by the geometrical parameters $h_{1}$ and $h_{2}$. Their importance
becomes evident as we explore the conditions under which the material
is $P\cdot k$ negative.

Maxwell's equations tell us that $P\cdot k=\frac{k_{X}^{2}}{\varepsilon_{Y}}+\frac{k_{Y}^{2}}{\varepsilon_{X}}$
in a uniaxial anisotropic material in $\mathbf{H}$ polarization.
Comparing to Eq. \ref{eq:permittivity_anis} we obtain $P\cdot k=k^{2}\mu_{\textrm{eff}}$.
The sufficient condition to have a $P\cdot k$ negative material in
$\mathbf{H}$ polarization is therefore that the effective permeability
be negative. If in addition we require $k_{X}$ and $k_{Y}$ to be
real (ignoring absorption) for all angles of incidence we must also
impose $\varepsilon_{X}<0$ and $\varepsilon_{Y}<0$. Writing it out
explicitly using Eq. \ref{eq:relsanis_detail} we obtain the
rather restrictive:\begin{equation}
\begin{array}{cc}
\frac{h_{2}}{h_{1}}<\left|\frac{\varepsilon_{2}}{\varepsilon_{1}}\right|<\frac{h_{1}}{h_{2}}\textrm{ when }\varepsilon_{1}<0\textrm{ and }\frac{h_{2}}{h_{1}}<\frac{h_{1}}{h_{2}}\\
\frac{h_{1}}{h_{2}}<\left|\frac{\varepsilon_{2}}{\varepsilon_{1}}\right|<\frac{h_{2}}{h_{1}}\textrm{ when }\varepsilon_{2}<0\textrm{ and }\frac{h_{1}}{h_{2}}<\frac{h_{2}}{h_{1}}.\end{array}\label{eq:Pknegconditions}\end{equation}
The crucial role played here by the purely geometrical quantity $h_{1}/h_{2}$
combines with the unboundedness of $\varepsilon_{Y}$ noted above
to explain the critical dependence of the behavior of the composite
on the relative spatial placement of its building blocks \cite{Woodley2005,Zharov2005}.
Moreover, these relations give, for the first time to our knowledge,
a precise geometrical recipe for the conditions under which a composite
will exhibit true negative index properties. We believe that, along
with Eqs. \ref{eq:permittivity_anis} from which they are derived,
they will become useful in the theoretical and especially the experimental
study of negative index composites. 

We now study the effect of losses on the effective medium. We assume
the media are passive ($b>0$), and that the imaginary parts are comparatively
small ($b\ll\left|a\right|$) so that we can ignore quadratic terms
in $b$ such as $b_{1}b_{2}$, etc. Rearranging \ref{eq:permittivity_anis}
we get\[
\begin{array}{cc}
\varepsilon_{X}=\left\langle a\right\rangle +i\left\langle b\right\rangle \\
\varepsilon_{Y}=\left\langle a^{-1}\right\rangle ^{-1}+i\frac{\left\langle ba^{-2}\right\rangle }{\left\langle a^{-1}\right\rangle ^{2}}.\end{array}\]

Let us take a closer look at $\varepsilon_{Y}$. The real part is
the harmonic mean of the real parts of the permittivities of the layers.
These permittivities have different signs, and therefore this quantity
is not bounded a priori. Let us call it $A$. Rewriting $\varepsilon_{Y}$
we get\begin{eqnarray}
 & \varepsilon_{Y}=A+i\left\langle \frac{b}{a^{2}}\right\rangle A^{2}\label{eq:epsXasymp}\end{eqnarray}

The unboundedness of the harmonic mean has a much stronger impact
on the imaginary part than on the real part. Eq.~\ref{eq:epsXasymp}
is only valid when $\textrm{max}(b_{1},b_{2})\ll A^{-1}$. Other cases
are of lesser interest as losses dominate.

We now validate our model as formulated in Eq. \ref{eq:permittivity_anis}
with numerical simulations. The transmission of the homogeneous anisotropic
slab was computed using a 1D transfer (T) matrix method, while for
the alternating single negative stack we chose a scattering (S) matrix
method for its numerical stability. In what follows all angles and
phase differences are measured in degrees while the scale parameter
$r$ is dimensionless.

\begin{figure}
\begin{center}\includegraphics[%
  scale=0.8]{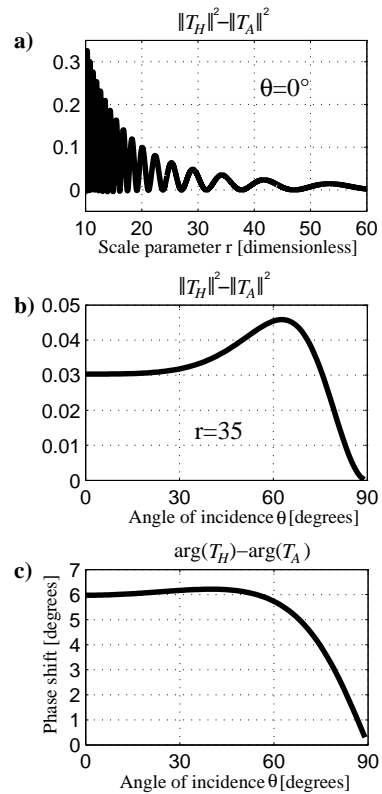}\end{center}

\caption{\label{cap:moduli-lambda} Offset between the power transmission
coefficients of single-negative stack ($\left\Vert T_{A}\right\Vert ^{2}$)
and the equivalent homogeneous medium ($\left\Vert T_{H}\right\Vert ^{2}$)
as a function of the scale parameter $r=\lambda/d$ for normal incidence
(a) and of angle of incidence for $r=35$ (b). They are within 0.05
of one another for $r>30$ for all angles of incidence. c) Phase mismatch
between the transmission coefficients of single negative stack and
effective medium as a function of incidence angle for $r=35$. They
are within 7° of each other for all angles of incidence.}
\end{figure}

The structure under study is composed of 200 alternating layers of
thicknesses $h_{1}=0.4$, $h_{2}=0.6$, with permittivities $\varepsilon_{1}=2$,
$\varepsilon_{2}=-2.8$ and permeabilities $\mu_{1}=-3.295$, $\mu_{2}=0.53$.
The equivalent parameters resulting from Eqs.~\ref{eq:permittivity_anis}
are $\varepsilon_{X}=-0.4$, $\varepsilon_{Y}=-70$ and $\mu_{\textrm{eff}}=-1$.
The total thickness of the structure is 100 periods. We now present
numerical simulations comparing the transmission coefficient of the
1D alternating layer structure, $T_{A}$, to that of the equivalent
homogeneous structure, $T_{H}$, as a function of $r$ and of angle
of incidence.

Figure 2.a) shows the offset between the power transmission coefficients,
$\left\Vert T_{H}\right\Vert ^{2}-\left\Vert T_{A}\right\Vert ^{2}$,
as a function of the scale parameter $r$ for normal incidence. The
same offset as a function of angle of incidence appears in Fig. 2.b)
for $r=35$. Finally, in Fig. 2.c) we plot the phase mismatch between
the two transmission coefficients as a function of angle of incidence
for $r=35$. These results demonstrate clearly that the meta-material
behaves in all respects as a homogeneous medium with the effective
parameters given by Eqs.~\ref{eq:permittivity_anis} for periods
more than 20 to 30 times smaller than the wavelength. 

The behavior at shorter wavelengths can be described by pushing the
developments of Eq. \ref{eq:multiscale} to higher orders. The
simple effective medium description is then no longer \emph{quantitatively}
valid but, interestingly, negative index-\emph{like} behaviors remain
\cite{Yoon2005}. More precise results in this direction will be presented
elsewhere.

We have this far detailed only the last of the three main consequences
of Eq. \ref{eq:permittivity_anis}. It is expressed quantitatively
in Eqs \ref{eq:relsanis_detail} and \ref{eq:Pknegconditions}.
We now discuss the first two: that negative index resonator/wire composites
are anisotropic and that their anisotropy can become extreme due to
the \emph{divergence of the harmonic mean}.

The significance of the anisotropy of the effective medium resides
in the difficulty of interpreting experimental data from a single
angle of incidence. The method of extracting effective parameters
that is most commonly used in the literature \cite{Smith2002,Greegor2003,Aydin2004,Zhang2005}
may fail when faced with an anisotropic material, particularly when
it is in addition indefinite. This aspect points to the need for a
new extraction procedure, one that can unambiguously place the sample
under study within one of the cases listed by Smith and Schurig \cite{Smith2003}
at a given frequency. Such a procedure would necessarily involve transmission
and phase measurements over \emph{multiple} angles of incidence.

The implications of the second point stretch beyond the anisotropy
of resonator/wire composites. Among other things, it tells us that
it is possible to realize materials with very large \emph{positive
or negative} permittivities or permeabilities using low valued single
negative stacks. It also enables a new understanding of enhanced transmission
through very thin slits in thick metal screens \cite{Porto1999} by
treating them as 1D air/metal stacks, providing a general theoretical
framework for the results of Shen et. al \cite{Shen2005}. Another
promising application is related to plasmon lasing \cite{Sirtori1998,Nezhad2004}.
Our preliminary results show that by stacking thin layers of active
material with thick metallic layers it may be possible to use the
divergence of the harmonic mean to dramatically enhance the gain of
the active material. More precise results and examples will be published
elsewhere.

In conclusion, we have argued that, because of the physical nature
of the negative permittivity phenomenon in metallic gratings, composites
employing them may be simply modeled as 1D single negative stacks.
Rows of wires make up effective negative permittivity slabs while
rows of resonators (metallic or dielectric) provide the negative permeability.
We have shown using a rigorous self-contained homogenization theory
that light sees the harmonic mean of the dielectric constant when
propagating parallel to the layers. Since the dielectric constant
is not positive definite the harmonic mean is not bounded. This remarkable
phenomenon, never before pointed out, explains quantitatively \emph{}the
sensitivity of the composites' behavior to their microscopic geometry
and provides clear criteria for constructing a true negative index
metamaterial. Furthermore, it tells us that it may now be possible
to fabricate media with arbitrarily large positive or negative effective
parameters using conceptually simple and technologically straightforward
1D single negative stacks. More in depth investigations of the possible
applications of this phenomenon are promising directions for further
research.

This work was supported by the French RNRT project CRISTEL and by
the EU-IST project PHAT 510162.

\bibliographystyle{apsrev}

\end{document}